\documentclass{vgtc}                          % final (conference style)
\graphicspath{{figures/}{pictures/}{images/}{./}} % where to search for the images

\usepackage{times}                     % we use Times as the main font
\usepackage{tabu}                      % only used for the table example
\usepackage{booktabs}                  % only used for the table example
\usepackage{lipsum}                    % used to generate placeholder text
\usepackage{mwe}                       % used to generate placeholder figures

\usepackage{mathptmx}                  % use matching math font

\usepackage{xcolor}

\usepackage{soul}
\definecolor{hlc}{HTML}{FFFFB3}  %e9e3f4
\sethlcolor{hlc}

\colorlet{framegray}{gray!30} % soft gray

\onlineid{0}

\vgtccategory{Research}

\vgtcinsertpkg

\title{Comparing the Emotional Impact of Thematic Versus Episodic
Framing in Visualization Text}

\author{Poorna Talkad Sukumar\thanks{e-mail: poorna.t.s@gmail.com}\\ %
        \parbox{2in}{\scriptsize \centering Department of Media Communications \\ Munster Technological University}  %
\and Maurizio Porfiri\thanks{e-mail: mporfiri@nyu.edu}\\ %
     \parbox{2in}{\scriptsize \centering Tandon School of Engineering \\ New York University} %
\and Oded Nov\thanks{e-mail: on272@nyu.edu}\\ %
     \parbox{2in}{\scriptsize \centering Tandon School of Engineering \\ New York University}}

\teaser{
    \vspace*{-0.1in}
  \centering
  \includegraphics[width=1\linewidth]{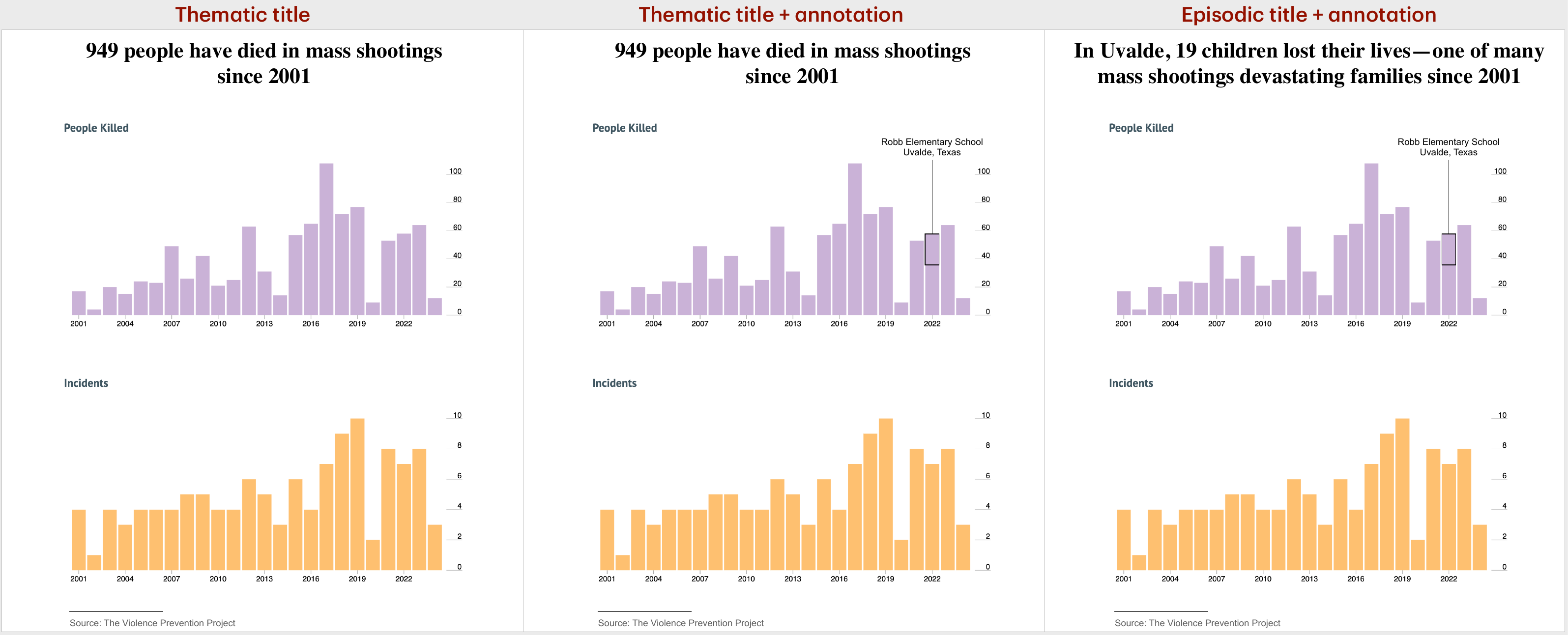}

  \vspace*{-0.1in}
  
  \caption{Experimental conditions showing textual framing manipulations. All conditions display identical bar charts of U.S. mass shooting data (2001–2024), with the top chart showing the number of people killed per year and the bottom chart showing the number of incidents per year; only the title and annotation vary. Left (T): a thematic title summarizing overall deaths (“949 people have died in mass shootings since 2001”). Center (T+Ann): the same title with an annotation marking a specific event (“Robb Elementary School, Uvalde, Texas”). Right (E+Ann): an episodic title referencing that event (“In Uvalde, 19 children lost their lives—one of many mass shootings devastating families since 2001”), paired with the same annotation. }
  \label{fig:teaser}
  
}

\abstract{
Although textual framing in data visualizations is known to influence comprehension, recall, and perceptions of bias, its effects on viewers' emotional responses remain underexplored. Drawing on two widely studied framing strategies in political communication, we examine how \textit{episodic} framing (foregrounding a specific event) versus \textit{thematic} framing (foregrounding broader trends) affects emotional and attitudinal responses to visualizations.
We conducted a preregistered, between-subjects online experiment (N = 800) in which participants viewed identical visualizations of U.S. mass shooting data that varied only in textual framing: a thematic title, a thematic title with annotation, or an episodic title paired with the same annotation. Results show that episodic framing elicited significantly more negative emotional valence than both thematic conditions. In contrast, adding an annotation to a thematic title did not alter emotional impact. While framing did not significantly affect policy attitudes, mediation analysis revealed a significant indirect effect: increased negative emotion under episodic framing predicted greater support for gun control.
These findings position emotion as a critical, yet underexamined, dimension of how textual framing shapes responses to data visualizations.
} % end of abstract


\begin{document}

%% The ``\maketitle'' command must be the first command after the
%% ``\begin{document}'' command. It prepares and prints the title block.

%% the only exception to this rule is the \firstsection command
\firstsection{Introduction}

\maketitle

Textual elements in data visualizations—such as titles, annotations, and captions—play a central role in shaping how viewers attend to and interpret visualizations. These elements guide attention and serve as anchors for interpretation, influencing what viewers remember and ultimately take away from a visualization \cite{borkin2015beyond, bylinskii2015eye, kim2021towards, stokes2022striking, zhu2022captions}.  A growing body of research has examined how textual framing in visualization influences comprehension, recall, and perceptions of bias or neutrality \cite{kong2018frames, kong2019trust, stokes2023role}, where framing refers to shaping how people interpret issues by emphasizing certain aspects while omitting others \cite{entman2007framing}. %Even subtle differences in title wording can shift how the same chart is understood, emphasizing certain aspects of the data while downplaying others.

However, textual framing is not just about cognitive resonance, but also about emotional processes \cite{aaroe2011investigating}. By selectively emphasizing certain aspects of a dataset, framing can shape not only what people think, but also how they \textit{feel} \cite{prantl2025untangling}. This emotional dimension of framing is particularly important in visualizations that communicate about contentious or socially significant issues, where emotional responses may influence interpretation, engagement, and attitudes \cite{sukumar2024connections, sukumareurovis}. 

Prior visualization research has shown that framing techniques such as proximity-based framing can influence viewers' emotional responses \cite{campbell2019feeling}. However, political communication and journalism studies identify a richer typology of framing strategies that remain underexplored in visualization contexts \cite{de2005news}. In this work, we focus on one such distinction: episodic versus thematic framing \cite{iyengar1994anyone}.
%While prior visualization research has largely examined framing in relatively simplified forms, political communication and journalism studies identify a richer typology of framing strategies that remain underexplored in visualization contexts \cite{de2005news}. In this work, we focus on one such distinction: episodic versus thematic framing \cite{iyengar1994anyone}. 
\textit{\textbf{Episodic}} frames foreground specific events or individuals, emphasizing concrete and vivid instances, whereas \textit{\textbf{thematic}} frames highlight broader patterns and trends. In text-based political communication, these two frame types are known to evoke systematically different emotional responses, and their relative persuasive force is conditioned by emotion \cite{aaroe2011investigating}. This distinction has not been examined in the context of visualizations and we investigate this gap in the domain of mass shooting visualizations—a high-stakes and emotionally charged setting where framing effects are particularly consequential for public discourse \cite{silva2019media, talkad2024are}.

We report a preregistered between-subjects online experiment (N = 800) in which all participants viewed identical bar charts of U.S. mass shooting data from 2001 to 2024, with number of deaths shown in the top chart and incidents in the bottom chart (see Figure \ref{fig:teaser}). Participants were randomly assigned to one of three conditions that varied only in textual framing: (1) a thematic title alone (T), (2) the same thematic title paired with an event annotation (T+Ann), or (3) an episodic title foregrounding a specific incident paired with the same annotation (E+Ann). We measured emotional valence and support for gun control policies both before and after exposure. This design allowed us to examine whether text-driven framing shifts emotional responses and attitudes, and whether emotional change mediates any observed attitude change. 

%We found that textual framing significantly influenced emotional responses but not policy attitudes directly. While all participants reported more negative emotional valence after viewing the visualization, this decrease was substantially larger in the episodic framing condition: episodic titles paired with annotations elicited significantly stronger negative emotional responses than both thematic conditions, whereas adding an annotation to a thematic title did not alter emotional impact. In contrast, framing did not significantly affect post-exposure support for gun control after controlling for baseline attitudes. However, mediation analysis revealed a significant indirect effect: episodic framing increased negative emotional responses, which in turn predicted greater support for gun control, despite the absence of a direct effect of framing on attitudes. Together, these results indicate that episodic framing operates primarily through affect, shaping attitudes indirectly rather than through direct persuasion.

We found that textual framing significantly influenced emotional responses but not policy attitudes directly. While all participants reported more negative emotional valence after viewing the visualization, this decrease was significantly larger under episodic framing; in contrast, adding an annotation to a thematic title did not significantly alter emotional impact. Although framing did not significantly affect post-exposure support for gun control after controlling for baseline attitudes, mediation analysis revealed a significant indirect effect: increased negative emotion under episodic framing predicted greater support for gun control, indicating that framing operates primarily through affect rather than direct persuasion.

%This work makes three key contributions. First, it establishes emotional response as a critical yet underexamined dimension of textual framing in visualization, demonstrating that titles and annotations influence not only interpretation but also affective experience. Second, it examines episodic versus thematic framing within visualization research, providing empirical evidence that these framing strategies differ in their emotional impact. Third, it identifies an emotion-mediated pathway linking framing to attitudes, showing that visualizations can shape public opinion indirectly through affect even when direct persuasive effects are absent. Together, these contributions advance our understanding of textual framing as a powerful rhetorical mechanism in communicative visualizations and highlight the importance of considering emotional as well as cognitive effects in visualization design.

This work makes three key contributions. First, it establishes emotional response as a critical, yet underexamined, dimension of textual framing in visualization, showing that titles and annotations shape not only interpretation but also affect. Second, it examines episodic versus thematic framing in visualization, providing empirical evidence that these strategies differ in their emotional impact. Third, it identifies an emotion-mediated pathway linking framing to attitudes, demonstrating that visualizations can influence public opinion indirectly through affect even in the absence of direct persuasive effects. Together, these contributions advance our understanding of textual framing as a rhetorical mechanism in communicative visualizations and highlight the importance of considering both emotional and cognitive effects in visualization design.

\section{Background}
\subsection{Text in Data Visualization}

Recent visualization research has established that textual elements are not peripheral to charts—they are central to how readers interpret them. Eye-tracking studies have shown that viewers often fixate first on text elements such as titles, legends, and captions, using them as anchors to guide interpretation before examining the underlying data \cite{bylinskii2015eye}. These elements influence what readers remember and take away from a visualization: titles that emphasize particular aspects of the data can direct attention toward some features while downplaying others \cite{borkin2015beyond, kong2018frames}. 

Annotations further extend this role by connecting visualized data to contextual or real-world meaning. Prior work distinguishes between observational text, which describes features of the data, and additive text, which introduces external information \cite{rahman2025survey}. Event annotations—such as labeling a specific incident within a trend—fall into this latter category, embedding real-world context directly within the visualization. Frameworks of semantic levels in visualization text position such contextual annotations as particularly influential, as they introduce external knowledge that can reshape interpretation \cite{lundgard2021accessible}. Indeed, viewers often prefer text-rich visualizations, valuing the interpretive guidance that annotations provide even at the cost of increased visual complexity \cite{stokes2022striking}.

Recent work recognizes that text in visualizations can convey affective meaning, for instance through “valenced subtext” that introduces emotionally charged or value-laden language \cite{stokes2025analysis}. However, while such work highlights the presence of emotional framing in visualization text, its impact on viewers’ emotional responses remains largely unexamined.

\subsection{Episodic and Thematic Framing}

%Framing theory describes how the presentation of information shapes interpretation by emphasizing certain aspects of an issue while omitting others \cite{entman2007framing}. 
A widely studied distinction in political communication is between episodic and thematic framing \cite{iyengar1994anyone}. 
Episodic frames typically elicit stronger emotional reactions than thematic frames \cite{aaroe2011investigating, gross2008framing}. Additionally, their persuasive effectiveness is shaped by emotion: episodic frames gain persuasive strength as emotional arousal intensifies, because they provide a specific focal point at which emotional reactions can be directed, whereas thematic frames—lacking this focal point—rely more on statistical and rational appeals \cite{aaroe2011investigating, gross2008framing}. 
We draw on this body of work to inform our hypotheses.

Although episodic and thematic framing are often presented as distinct categories, they are better understood as points along a continuum. Communication artifacts frequently contain both episodic and thematic elements to varying degrees. In this work, we examine one operationalization of episodic and thematic framing by manipulating the accompanying textual framing while holding the visualization design constant.

\section{Experiment}

%We conducted a preregistered, between-subjects online experiment to examine how episodic versus thematic textual framing in data visualizations influences emotional responses and attitudes toward gun control.
%Participants were randomly assigned to one of three conditions (see Figure \ref{fig:teaser}) showing identical visualizations of U.S. mass shooting data (2001–2024), differing only in textual elements: a thematic title (T), a thematic title with an event annotation (T+Ann), or an episodic title foregrounding the same event paired with the annotation (E+Ann).
%We measured emotional valence and support for gun control both before and after exposure. This design enabled us to assess framing effects on both emotional change and attitude change, as well as to test whether emotional responses mediate attitudinal effects.

%We set up our study as an online survey on Qualtrics and we tested the following  hypotheses:\\
%\textbf{H1}: Episodic framing will produce more negative emotional change than thematic framing (E+Ann $<$ T+Ann $<$ T). \\
%\textbf{H2:} Episodic framing will lead to greater increases in support for gun control (E+Ann $>$ T+Ann $>$ T). \\
%\textbf{H3:} Changes in emotional valence will mediate the relationship between framing condition and policy support.

We set up our study as an online survey on Qualtrics and tested the following hypotheses:\\
\textbf{H1:} Episodic framing will produce more negative emotional change than thematic framing, with emotional change expected to be most negative under episodic framing, followed by thematic with annotation, and least under thematic alone (E+Ann $<$ T+Ann $<$ T). \\
\textbf{H2:} Episodic framing will lead to greater increases in support for gun control than thematic framing. Accordingly, we expect policy support to be highest under episodic framing, followed by thematic framing with annotation, and lowest under thematic framing alone (E+Ann $>$ T+Ann $>$ T). \\
\textbf{H3:} Changes in emotional valence will mediate the relationship between framing condition and policy support.

Additional analyses included manipulation checks (perceived framing and policy direction and annotation detection), and baseline equivalence.   %, and exploratory moderation by political ideology.

\vspace*{0.5em}

\noindent\textbf{Participants:} We recruited 800 participants through the Prolific platform. Participants were required to be at least 18 years old, be fluent in English,
be located in the United States, and complete the survey on a desktop or laptop device. The median completion time across all experimental conditions was approximately six minutes, and the participants were paid \$1.50 for their participation. The target sample size was preregistered based on an a priori power analysis and increased to account for anticipated exclusions and planned exploratory analyses.

\vspace*{0.5em}

\noindent\textbf{Study Material:} All participants viewed the same underlying visualization, consisting of two vertically stacked bar charts: the top chart shows the number of people killed, and the bottom chart shows the number of incidents from 2001 to 2024 (see Figure \ref{fig:teaser}). The visualization design, including the use of vertically stacked charts and annotation style, was inspired by similar examples in data journalism, particularly those published by \textit{The New York Times} \cite{nytimes}. Data for the conditions were obtained from the Violence Project Mass Shooter Database \cite{violenceproject}, with the source indicated in the visualization across all conditions. %All the conditions were static and are included in the supplementary material.

Across conditions, the charts were identical in terms of data, axes, colors, and layout. Only the textual elements (title and annotation) were manipulated to vary framing. For the thematic condition, we drew on common headline styles in news articles that emphasize aggregate trends \cite{washpost} to construct the title “949 people have died in mass shootings since 2001.”

For episodic framing, we selected the Robb Elementary School shooting in Uvalde, Texas (2022) as the focal event. This event was chosen because it is recent, widely recognized, and represents a salient instance within the dataset, frequently highlighted in media coverage \cite{washpost}. The episodic title foregrounded this event while situating it within the broader trend (“In Uvalde, 19 children lost their lives—one of many mass shootings devastating families since 2001.”)

%Because episodic framing typically foregrounds a specific event, it is often accompanied by an explicit callout to that event within the visualization. Accordingly, we paired the episodic title with an annotation marking the same incident (“Robb Elementary School, Uvalde, Texas”). We included an additional condition pairing the thematic title with the same annotation to examine whether adding an annotation changes the effect of a thematic title. 

Because the episodic title explicitly foregrounds a specific event, we paired it with an annotation labeling that event (``Robb Elementary School, Uvalde, Texas'') to create an ecologically realistic instantiation of episodic framing. Without such a reference, readers might expect to locate the highlighted event in the chart. Our goal was to compare a realistic episodic framing strategy against thematic alternatives. We therefore also included a thematic-title-plus-annotation condition to assess whether any observed effects were attributable to the annotation itself.
We acknowledge that a fully crossed design including an episodic-title-without-annotation condition would allow a cleaner separation of title and annotation effects and represents an important direction for future work.

\vspace*{0.5em}

\noindent\textbf{Survey Procedure: }
Participants first provided informed consent and then completed pre-exposure measures of emotional valence (using the 9-point Self-Assessment Manikin [SAM] scale \cite{lang2005international}) and support for gun control policies. Support for gun control was assessed by asking whether firearm laws should be made stricter, rated on a 5-point Likert scale ranging from “Strongly disagree” to “Strongly agree”.

Participants were then randomly assigned to one of the three experimental conditions and were required to view the visualization for at least 15 seconds before proceeding. To ensure engagement, we included two factual questions based on the visualization’s content.
After viewing the visualization, participants completed post-exposure measures of emotional valence and policy attitudes. This was followed by manipulation checks assessing perceived framing (episodic vs. thematic) and whether participants noticed the annotated event, as well as a verification question to confirm recognition of the topic. Finally, participants provided demographic information, including age, gender, education, and political ideology, and were debriefed. %The Qualtrics survey is included in the supplementary material.

\section{Results}

We excluded participants who incorrectly answered the topic verification question, as well as those that failed to answer both the factual
questions. In total, 17 of the 800 participants were removed. There were 263, 259, and 261 participants in the T, T+Ann, and E+Ann conditions, respectively.

\subsection{Manipulation Checks}

Manipulation checks were included to ensure that the framing manipulations were interpreted by participants as intended: \\
\textbf{Perceived framing.} Participants perceived the framing hierarchy as intended, with episodic emphasis increasing from T to T+Ann to E+Ann (linear trend estimate = 1.89, t(780) = 11.50, p $<$ .001). \\
\textbf{Annotation noticed.} Participants were substantially more likely to report noticing an annotation in annotation-present conditions (odds ratios $>$ 400, ps $<$ .001), confirming that the annotation manipulation was highly salient. \\
\textbf{Perceived policy direction.} Although we expected no differences in perceived policy direction (pro-gun) across conditions, a significant effect was observed, F(2, 780) = 7.26, p $<$ .001, with the episodic condition perceived as more pro–gun control than the thematic condition.\\
\textbf{Baseline equivalence.} There were no significant differences in pre-exposure valence or gun-control support across conditions.

\begin{figure}[!t]
  \centering
  \includegraphics[width=1\linewidth]{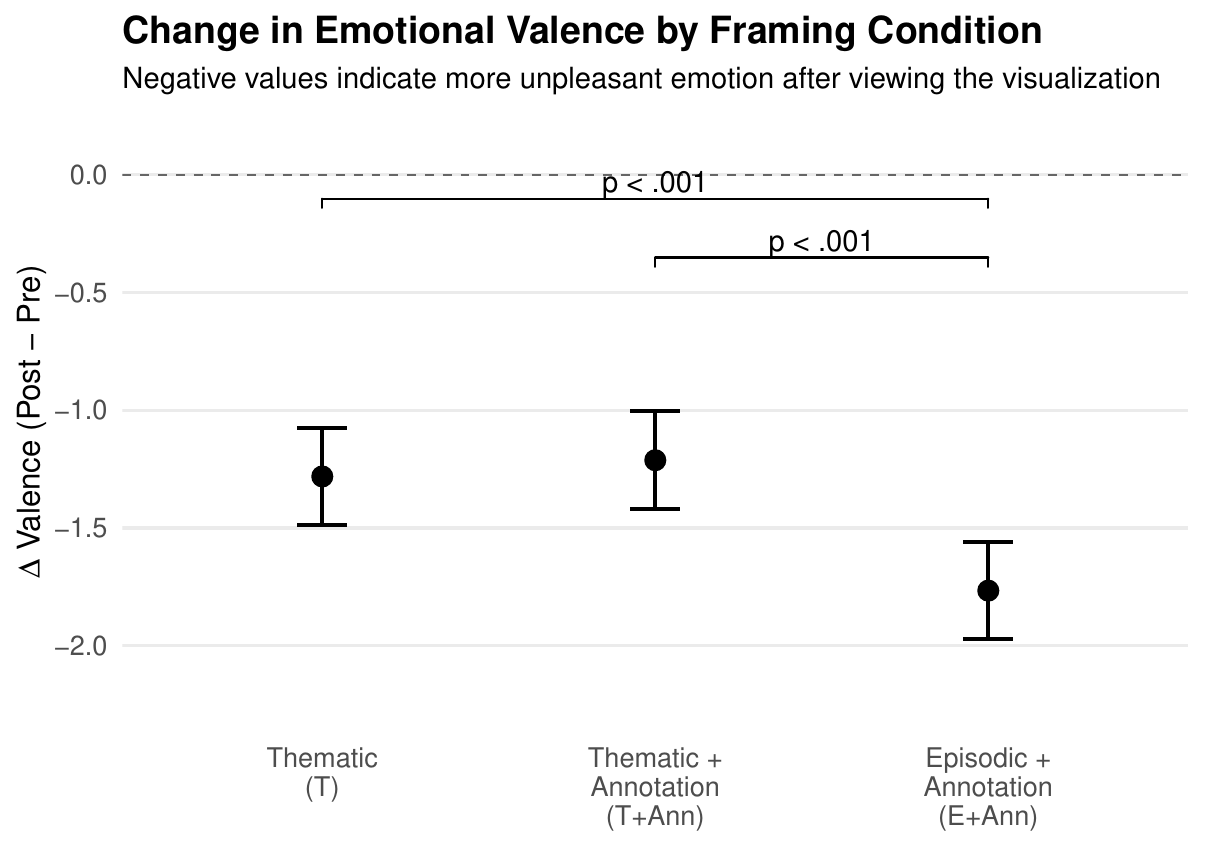}
    \vspace*{-0.25in}
 
\caption{Estimated change in emotional valence by framing condition. Points show estimated pre--post change in valence, with 95\% confidence intervals. Negative values indicate more unpleasant emotional responses after viewing the visualization. Significant planned contrasts between conditions are annotated. }
  \label{means}
  
    \vspace*{-\baselineskip}
\end{figure}

\subsection{H1: Emotional Response (Valence Change)}

To test H1, we fit a linear mixed-effects model predicting valence with Time (Pre vs.\ Post) as a within-subject factor and Condition (T, T+Ann, E+Ann) as a between-subject factor, with a random intercept for participant.

There was a significant main effect of Time, $F(1, 780) = 546.52$, $p < .001$, indicating that participants reported lower (more negative) valence after viewing the visualization. Crucially, the Time $\times$ Condition interaction was also significant, $F(2, 780) = 8.22$, $p < .001$, indicating that the magnitude of emotional change differed across framing conditions (see Figure \ref{means}).

Within-condition contrasts showed that all conditions produced a significant decrease in valence (all $p$s $< .001$):
\begin{itemize}\setlength{\itemsep}{0pt}
    \item T: $\Delta = -1.28$, 95\% CI [$-1.49$, $-1.08$]
    \item T+Ann: $\Delta = -1.21$, 95\% CI [$-1.42$, $-1.01$]
    \item E+Ann: $\Delta = -1.77$, 95\% CI [$-1.97$, $-1.56$]
\end{itemize}

Planned contrasts on change scores showed that E+Ann produced a significantly larger decrease in valence than both T and T+Ann (both $p$s $< .001$), while no difference was observed between T and T+Ann ($p = .64$).

These results indicate that episodic framing produced significantly stronger negative emotional responses than thematic framing, while adding an annotation to a thematic title had no measurable effect. Thus, H1 is partially supported.

\subsection{H2: Attitude Change (Support for Gun Control)}

To test H2, we fit an ANCOVA predicting post-exposure support with Condition as a factor and pre-support as a covariate, with ideology included as a control. There was no significant effect of Condition, $F(2, 771) = 1.46$, $p = .23$.

Adjusted means were nearly identical across conditions:
\begin{itemize}\setlength{\itemsep}{0pt}
    \item T: $4.06$ [4.02, 4.11]
    \item T+Ann: $4.06$ [4.01, 4.10]
    \item E+Ann: $4.11$ [4.06, 4.15]
\end{itemize}

\noindent Planned contrasts were non-significant (all $p$s $> .10$). These results indicate that framing did not significantly influence policy attitudes after controlling for baseline support. Hence H2 is not supported.

\subsection{H3: Mediation via Emotional Response}

We tested whether emotional change mediated the relationship between framing and policy attitudes using a mediation model with bootstrapped standard errors (5{,}000 samples).

Episodic framing (E+Ann vs. T+Ann) significantly increased negative emotional change ($a_2 = -0.34$, $p < .001$), and greater negative emotional change was associated with higher post-exposure support for gun control after controlling for baseline support and political ideology ($b = -0.04$, $p < .001$). The indirect effect was significant ($\mathrm{ind}_2 = 0.013$, $p = .011$), while the direct effect was not ($p = .36$).

%This finding was robust when modeling change in support ($\Delta support=post - pre$) directly: greater negative emotional change remained associated with greater increases in support for gun control ($b = -0.033$, $p = .001$), and the indirect effect remained significant ($\mathrm{ind}_2 = 0.011$, $p = .019$).

In contrast, the annotation-only contrast (T+Ann vs.\ T) showed no significant effects on emotional change or mediation pathways.
These results indicate that episodic framing influenced policy attitudes indirectly through increased negative emotion, despite the absence of a direct effect. Hence H3 is supported.

\section{Discussion and Design Implications }

\subsection{Emotional Effects of Episodic Framing}

The most robust finding of our study is that episodic framing produced significantly stronger negative emotional responses than thematic framing. Participants in the episodic condition (E+Ann) exhibited a substantially larger decrease in valence compared to both thematic conditions, while the addition of an annotation alone (T+Ann) did not alter emotional responses.

These findings are consistent with theories of episodic and thematic framing \cite{iyengar1994anyone}. By foregrounding a specific event or individual, episodic framing may anchor attention to a concrete referent, intensifying emotional responses relative to trend-focused framing.

Finally, adding an event annotation to a thematic title did not significantly increase emotional responses, suggesting that simply labeling a specific event was insufficient to amplify emotion. One possible explanation is that the annotation may assume some familiarity with the event, whereas the episodic title explicitly states that 19 children lost their lives, communicating the emotional significance directly even to readers without prior knowledge. These findings are consistent with prior work suggesting that titles play an important role in establishing the interpretive frame of a visualization \cite{borkin2015beyond,kong2018frames,kong2019trust}.

\noindent\textbf{Design Implication 1. Use episodic framing to increase emotional engagement, but recognize its editorial tradeoffs.}
Episodic framing may make abstract, large-scale data more emotionally accessible by grounding it in a concrete human experience. However, media coverage of mass shootings disproportionately emphasizes a small number of highly salient events, shaping public understanding at the expense of less prominent cases \cite{silva2019media}. Consequently, foregrounding an event such as Uvalde represents an editorial choice about which incidents become emotional anchors for interpreting the broader dataset.

\vspace*{-0.5em}
\subsection{Emotion Without Direct Persuasion}

Despite strong emotional effects, we did not observe a direct effect of framing on policy attitudes. This result is consistent with prior work suggesting that deeply held attitudes—particularly on politically charged issues—are often resistant to change following a single visualization exposure \cite{heyer2020pushing, kong2018frames, pandey2014persuasive, sukumar2024connections}.

However, mediation analysis revealed a more nuanced pattern. %Greater negative emotional change predicted increased support for gun control, suggesting that textual framing may influence attitudes indirectly by first shaping viewers' emotional responses. 
Greater negative emotional change was associated with greater increases in support for gun control, suggesting that textual framing may influence attitudes indirectly by first shaping viewers' emotional responses.

\noindent\textbf{Design Implication 2. Emotional engagement alone may not be sufficient for immediate attitude change.}
Designers seeking to influence attitudes over time may  need to more explicitly connect the emotional experience elicited by a visualization to its broader societal or policy implications.

\subsection{Effects of Perceived Policy Direction}

An interesting finding was that participants perceived the episodic condition as more pro–gun control, despite all conditions presenting the same data and a broadly pro–gun control framing.

Prior work has shown that textual elements in visualizations can strongly influence perceptions of author bias and the neutrality of the data \cite{stokes2023role}. Extending this work, our results suggest that emphasizing a specific event can also shift readers' perceptions of the communicator's policy stance, even when the underlying visualization and data remain unchanged.

\noindent\textbf{Design Implication 3. Textual framing can shape perceived editorial stance independently of the data.}
Designers working in contexts where credibility and perceived neutrality are important should evaluate textual framing alongside visualization design. Choices about which events are foregrounded contribute to the rhetorical message of a visualization and may influence readers' perceptions of advocacy and editorial intent.

\section{Limitations and Future Work}

This study focuses on a single domain—mass shootings in the United States—which is highly salient and emotionally charged. While appropriate for studying framing effects, future work should examine whether these findings generalize to other domains.

Our manipulation represents one operationalization of episodic and thematic framing. For example, an even more episodic title could identify a specific victim by name and personal details. Likewise, our visualization itself presents aggregate counts over time, reflecting a primarily thematic visual structure. More generally, visualization design can also shape episodic and thematic interpretations—for example, Periscopic's \textit{U.S. Gun Deaths} visualization \cite{periscopic} transitions from individual victims to an aggregate overview—while prior work has shown that design choices such as aggregation level influence reasoning about individual cases versus broader population-level patterns \cite{holder2022dispersion, xiong2019illusion}. Future work should investigate how textual and visual framing jointly shape viewers' emotions and interpretations.

We also examine a specific form of annotation inspired by journalistic practice—event-based annotations highlighting a particular incident and its location (e.g., \cite{nytimes}). This represents only a small subset of the broader annotation design space, which includes textual, graphical, and hybrid forms varying in function, placement, and level of detail~\cite{rahman2025survey}. Future work should investigate how different annotation types interact with framing to shape emotional and cognitive responses.

Finally, political communication research identifies a much broader range of framing strategies \cite{de2005news}. Exploring how other types of frames operate in visualization contexts may provide a useful direction for future research.

\section{Conclusion}

%We investigated how episodic versus thematic framing in visualization titles and annotations shapes emotional responses and policy attitudes toward gun control. Across 800 participants viewing identical charts, episodic framing produced significantly stronger negative emotional responses than thematic framing, while the addition of an annotation alone had no measurable affective impact. Framing did not directly shift policy attitudes, but mediation analysis revealed a significant indirect pathway: episodic framing increased negative emotion, which in turn predicted greater support for gun control.  The episodic–thematic distinction offers an actionable lens for visualization design. While these framing choices are already common in visualization design practice, our results show that they have measurable consequences for how audiences feel about the data and, indirectly, what they support.

This study demonstrates that textual framing in data visualizations shapes not only how people interpret data, but also how they feel about it. Episodic framing amplifies emotional responses, and these emotional shifts can indirectly influence attitudes even in the absence of direct persuasion. These results may seem unsurprising—foregrounding 19 children killed in Uvalde is inherently more affecting than presenting an aggregate death toll. But our results reveal tensions that visualization designers and data journalists must navigate: episodic framing appears to drive emotional engagement at the cost of perceived neutrality, with participants reading identical data as more advocacy-oriented when framed episodically. By anchoring emotional responses to a specific event, it also risks privileging certain narratives over others. The choice between episodic and thematic framing is not merely stylistic—it carries measurable consequences for how audiences feel, what they believe the communicator intends, and whose stories get told.

\section{Preregistration and Supplementary Materials}

The preregistration for the study can be found at \url{https://osf.io/mcy7h/overview} and the supplementary materials can be found at \url{https://osf.io/fe9k6/}.

%% if specified like this the section will be committed in review mode
\acknowledgments{
We are very thankful to the reviewers for
providing valuable feedback. This work was supported by the National Science Foundation under Grant CMMI-1953135.}

\bibliographystyle{abbrv-doi}

\bibliography{template}
\end{document}